\documentclass{emulateapj} 

\usepackage{ifthen}
\newboolean{color}

\setboolean{color}{false}
\setboolean{color}{true}

\slugcomment{\\Submitted to ApJ}

\usepackage{amsmath}
\usepackage{graphicx}
\usepackage{color}
\usepackage{float}
\usepackage{amssymb,amsmath}
\usepackage{epsfig,floatflt}
\usepackage{subfigure}
\begin{document}

\newcommand \thi{\thinspace}
\newcommand \noi{\noindent}

\long\def\jumpover#1{{}}

\def\ni{{\noindent}}
\def \at{{\rm\char'100}}
\newcommand \eg{{{\it e.g.}\ }}
\newcommand \etal{{\it et al.\ }}
\newcommand \etc{{\it etc.\ }}
\newcommand \cf{{\it cf.\ }}
\newcommand \ia{{{\it inter alia}\ }}
\newcommand \ie{{{\it i.e.}\ }}
\newcommand \via{{\it via\ }}
\newcommand \adhoc{{\it ad hoc\ }}
\newcommand \viz{{\it viz.\ }}
\newcommand \vs{{\it vs.\ }}

\newcommand \Teff{{$T_{\rm {ef\!f}} $}}
\newcommand \teff{{T_{\rm {ef\!f}} }}
\newcommand \Lo{{$L_\odot $}}
\newcommand \Mo{{$M_\odot $}}
\newcommand \Log{{\rm Log}\thi}
\newcommand \approxgt{\,\raise2pt \hbox{$>$}\kern-8pt\lower2.pt\hbox{$\sim$}\,}
\newcommand \approxlt{\,\raise2pt \hbox{$<$}\kern-8pt\lower2.pt\hbox{$\sim$}\,}
\newcommand\dotd{\hbox{$.\!\!^{\rm d}$}}
\newcommand\dotm{\hbox{$.\!\!^{\rm m}$}}
\newcommand\dd{{$^d$}}
\newcommand\ML{{$M$-$L$\ }}
\newcommand\aanda{A\&A}
\newcommand\RV{$V_r$}
\newcommand\DP{{$\Delta(\Delta\Phi_1)$}}
\newcommand\DDP{{$|\Delta(\Delta\Phi_1)|$}}
\newcommand\dinv {{\thinspace d$^{-1}$}}

\newcommand\MB{{M$_{\rm B}$}}
\newcommand\MR{{M$_{\rm R}$}}
\newcommand\OI{{OGLE I}}
\newcommand\OV{{OGLE V}}

\title{On the Blazhko Effect in RR~Lyrae Stars}
\author{J. Robert Buchler\altaffilmark{1} \&
Zolt\'an Koll\'ath\altaffilmark{2}}
\altaffiltext{1}{\ni Physics Department, University of Florida,
Gainesville, FL 32611, USA; buchler\at phys.ufl.edu}
\altaffiltext{2}{\ni Konkoly Observatory, Budapest, HUNGARY;
kollath\at konkoly.hu}

\begin{abstract}

The Blazhko effect is a long term, generally irregular modulation of the light
curves that occurs in a sizeable number of RR~Lyrae stars.  The physical origin
of the effect has been a puzzle ever since its discovery over a hundred years
ago.  We build here upon the recent observational and theoretical work of
\citet{setal} on RRab stars who found with hydrodynamical simulations that the
fundamental pulsation mode can get destabilized by a $9:2$ resonant interaction
with the 9th overtone.  Alternating pulsation cycles arise, although these
remain periodic, \ie not modulated as in the observations.  \\
Here we use the
amplitude equation formalism to study this nonlinear, resonant interaction
between the two modes.  We show that not only does the fundamental pulsation
mode break up into a period two cycle through the nonlinear, resonant
interaction with the overtone, but that the amplitudes are modulated, and that
in a broad range of parameters the modulations are irregular as in the
observations.  This irregular behavior is in fact chaotic and arises from a
strange attractor in the dynamics.

\end{abstract}


%%%%%%%%%%%%%%%%%%%%%%%%%%%%%%%%%%%%%%%%%%%%%%%%%%%%%%%%%%%%%%%%%%%%
\section{Introduction}

A large subclass of RR~Lyrae stars undergo light curve modulations, most of
them on the time scale of some 60 periods, although the range extends from some
tens to some hundreds of periods.  The fundamental pulsation cycle itself lasts
~$\sim$~0.5\thi day.  The effect was discovered by \citet{blazhko} over a
hundred years ago, and a number of explanations have been proposed, such as
closely spaced pulsation modes, a modal $1:2$ resonance, an oblique rotator
model, a nonradial modal interaction, and convective cycles
\citep{stothers,molnark}.  However, none of these mechanisms is without fault.

The giant step toward the explanation of the Blazhko effect has come from the
unprecedently precise and continuous {\sl Kepler} space telescope observations
and their analysis \citep{setal}.  Because we are going to build on this work
we first present a summary of these findings.  Fourteen RR Lyr stars in their
sample undergo Blazhko modulations.  Unexpectedly, three of these stars also
display period doublings, \ie the shapes of the light curves shows cycle to
cycle alternations.  The depths of these alternations change during the Blazhko
cycle. Another recent observational finding is that the Blazhko cycle does not
repeat regularly, see \eg \cite{chadid,kolenberg,sodor}.  This behaviour poses
another important constraint for the physical explanation of the Blazhko
effect, as do the observed variations of the mean physical parameters of the
stars during the Blazhko cycle \citep{jurcsik1,jurcsik2}.

On theoretical side \citet{setal} performed a systematic numerical hydrodynamical
modeling survey of RR~Lyr models.  They found that over a relatively broad
region of astrophysical model parameters the fundamental pulsation is unstable
and develops into a pulsation with alternating cycles.  Guided by earlier work
\citep{mb} that had shown that half-integer resonances can cause a bifurcation
to alternating cycles, \citet{setal} did some sleuthing work.  By computing the
Floquet stability coefficients \citep{hartman, bmk} of the fundamental
pulsation and searching for resonances of the fundamental mode with successive
overtones showed that it is the 9th overtone that is in a $9:2$ resonance and
destabilizes the fundamental pulsation cycle.  (It is of course a coincidence
that it should be the 9th overtone that is in a $9:2$ resonance.)  This 9th
overtone in these RR~Lyr models turns out to be egregious in that it is a
surface mode (dubbed 'strange mode' when it was first encountered by
\citet{strange1} and \citet{strange2}).  The $9:2$ resonance appears in a
relatively narrow, winding band in an $\Log L- \Log T$ diagram \citep{KMS}.

%-----------------------------------------------------------------------
\begin{figure*}[ht!]
\begin{center}
  \includegraphics[scale=0.9]{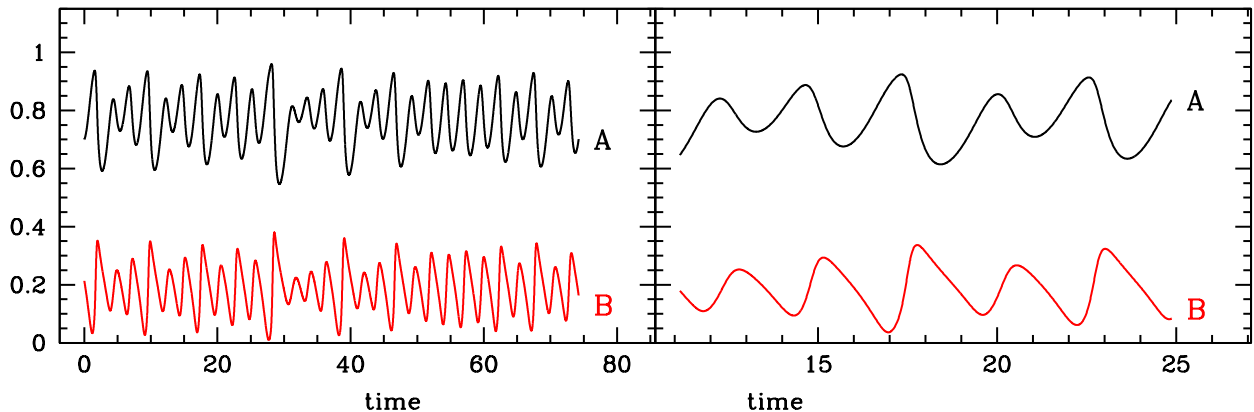}
\caption{Temporal modulation of the two amplitudes of the
irregularly  modulated period two  pulsation.}
\end{center}
\label{fig1}
\end{figure*}
%-----------------------------------------------------------------------

The numerical hydrodynamical simulations \citep{KMS} were able to produce
alternating cycles, and the ancillary analyses established that the $9:2$
resonance is the cause of the symmetry breaking bifurcation.  However, these
hydrodynamical simulations were unable to produce either regular or irregular
Blazhko like modulations.

When a surface mode plays a dynamical role, the numerical hydrodynamical
simulations become particularly sensitive to the mixing length parameters, to
the zoning in the outer regions as well as to the surface boundary condition.
It is therefore extremely difficult to make completely trustworthy and robust
simulations.  In fact it is an open question whether hydrodynamics with a time
dependent mixing length treatment of convection is able to produce Blazhko
modulations.  For these reasons it is of importance to resort to an
alternative, complementary approach as we do in this paper.

\section{Modeling}

In the 1980's several groups developed the amplitude equation formalism in an
effort to understand the systematics of stellar pulsations throughout the $HR$
diagram \citep[for a review, see \eg][]{b93}.  The validity of this powerful
formalism rests on the fact that for mildly nonadiabatic stars, such as the
RR~Lyrae and the classical Cepheids, one can decouple the 'fast' pulsation and
obtain equations for the slow secular behavior of the amplitudes and phases,
akin to the Poincar\'e - Lindstedt approach in celestial mechanics.  

The amplitude equation formalism is complementary to numerical hydrodynamical
simulations.  It allows one to understand the mathematical structure of the
dynamics of the stellar models.  As such it is therefore ideally suited for
making surveys of the possible pulsational behavior of a given type of star.
However, when light curves or radial velocity curves for specific models are
desired, one needs to resort to numerical hydrodynamics.

The form of these amplitude equations is generic and depends only on the
resonances that exist between the excited linear modes of pulsation
\citep[\eg][]{b93,spiegel}.  Amplitude equations have a solid mathematical
foundation and are also known as 'normal forms' in the mathematical literature
\citep{guck}.

We note in passing that the first success of the amplitude equation formalism
in stellar pulsation theory was to prove the conjecture of \citet{simon} that
the Hertzsprung progression of the Fourier phases of the bump Cepheid light
curves and radial velocity curves is indeed caused by the $2:1$ resonance
between the fundamental pulsation mode and the linearly stable second overtone
\citep{kgb,bump}.  We also note that \citet{skb} were able to map out
efficiently the pulsational behavior of RR Lyr stars throughout the whole
instability strip by using a judicious combination of numerical hydrodynamical
simulations and amplitude equations.

\vskip 5pt

In this paper we ignore first overtone (RRc) and double mode pulsations (RRd)
that we addressed in \eg \cite{kbsc,skb}.  We therefore do not need to include
an amplitude equation corresponding to the first overtone in what follows.
Hereafter, we label the fundamental mode with 'a' and the 9th overtone with 'b'.

The amplitude equations that are appropriate for the modal interaction between
mode a that is in a $9:2$ resonance with mode b ($9\omega_a \sim 2\omega_b$)
have already been given in \citet{mb}.

\begin{eqnarray}
{da \over dt}      &=& \big(i \omega_a +\kappa_a  - Q_a |a|^2 - T_a |b|^2\big) a +
                    {\color{black}  c_a  a^{*8} b^2} \nonumber \\
\noalign{\vskip 4pt}
{db \over dt}      &=& \big( i \omega_b +\kappa_b - Q_b |b|^2 - T_b |a|^2 \big) b +
                 {\color{black}   c_b  a^9 b^*} \nonumber \\
&&
\label{eqs0}
\end{eqnarray}

By introducing real amplitudes, $A$ and $B$, and phases $\phi_a$ and $\phi_b$,
by defining the relative phase
$$\Gamma = 9 \phi_b -2 \phi_a$$ 
we can cast Eqs.~\ref{eqs0} into a set of 3 real equations

\begin{eqnarray}
{dA \over dt}      &=& \big(\kappa_a  - Q_a A^2 - T_a B^2\big) A +
                    {\color{black}  C_a  A^8 B^2 \cos(\Gamma+\delta_a)} \nonumber \\
\noalign{\vskip 4pt}
{dB \over dt}      &=& \big( \kappa_b - Q_b B^2 - T_b A^2 \big) B +
                 {\color{black}   C_b  A^9 B \cos(\Gamma-\delta_b)} \nonumber \\
\noalign{\vskip 4pt}
 {\color{black} {d\Gamma \over dt}} & {\color{black}=}&  {\color{black}2 \Delta
  - 9 C_a A^7 B^2 \sin(\Gamma+\delta_a) -2 C_b A^9
    \sin(\Gamma-\delta_b) } \nonumber \\
&&
\label{eqs}
\end{eqnarray}

The $\kappa$ denote the growth rates of the two modes.  The quadratic terms
$Q_a$, $Q_b$, $T_a$ and $T_b$ appear in the amplitude equations even in the
absence of any resonances.  Their imaginary parts do not have a significant
effect and convenience we have assumed that they are all real.  The quantities
$C_a$, $C_b$ and their phases $\delta_a$, $\delta_b$ describe the resonant
coupling, and $$\Delta = \omega_b -\frac92 \omega_a$$ is the off-resonance
parameter.

%-----------------------------------------------------------

\vskip 5pt

Two types of solutions exist depending on the parameters of the amplitude equations:

\vskip 5pt

\noi 1. \underbar{The fixed points},
($dA/dt=0 ,dB/dt=0, d\Gamma/dt=0$). 

\vskip 3pt

 Eqs.~(\ref{eqs}) have two possible fixed points:

\vskip 2pt

i. $A=0$ and $B=0$.  This fixed point corresponds to the trivial, nonpulsating
static model.  It is unstable when mode a is self-excited, $\kappa_a>0$, \ie
inside the fundamental instability strip.

\vskip 2pt

ii. A single mode fixed point with $A=A_0=\sqrt{\kappa_a/Q_a}$ and $B_0=0$.
This fixed point, when stable, corresponds to steady periodic pulsations in the
fundamental mode with constant amplitude.  In some regime of parameters
this solution is unstable to an excitation of mode b, in which case we have

\vskip 2pt

iii. A two-mode fixed point with $A=A_0$ and $B=B_0$ (and $\Gamma=\Gamma_0$).
Note that because of the $9:2$ resonance condition, it takes {\sl two}
fundamental periods for the oscillation to repeat. 
This is the fixed point that \cite{KMS} found in their numerical hydrodynamical
simulations and that gives rise to the period alternations.

Generally, the equations~(\ref{eqs}) are  complicated enough so that the values
of $A_0$ and $B_0$ must be obtained numerically.  

In some parameter regime this
fixed point can also be unstable, as we describe now.

\vskip 5pt

%-----------------------------------------------------------------------
\begin{figure}[h!]
\begin{center}
  \includegraphics[scale=0.5]{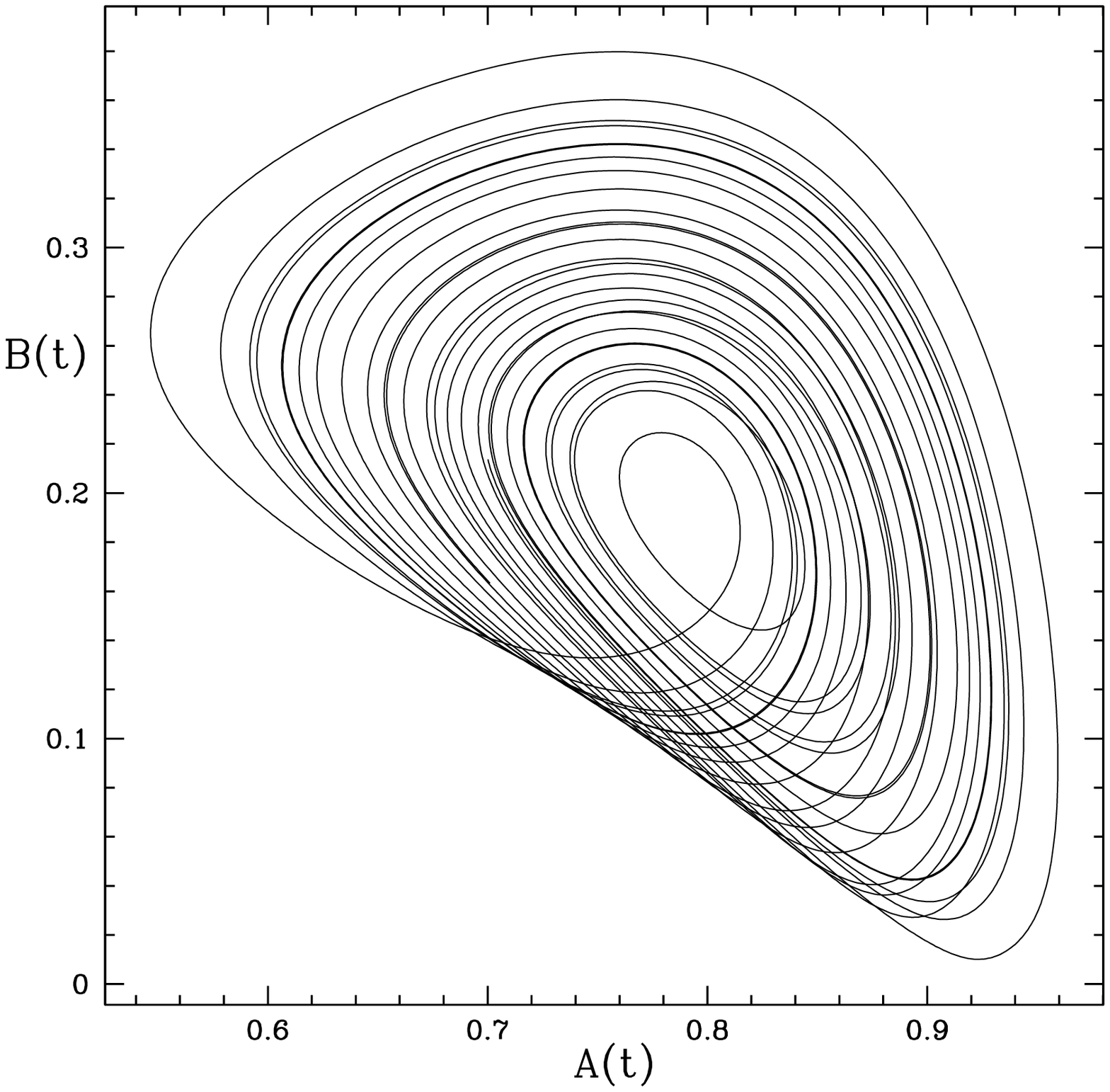}
\caption{Phase plot: amplitude B(t) versus amplitude A(t).}
\end{center}
\label{fig2}
%\end{figure}
%-----------------------------------------------------------------------

%-----------------------------------------------------------------------
%\begin{figure}[h!]
\begin{center}
  \includegraphics[scale=0.5]{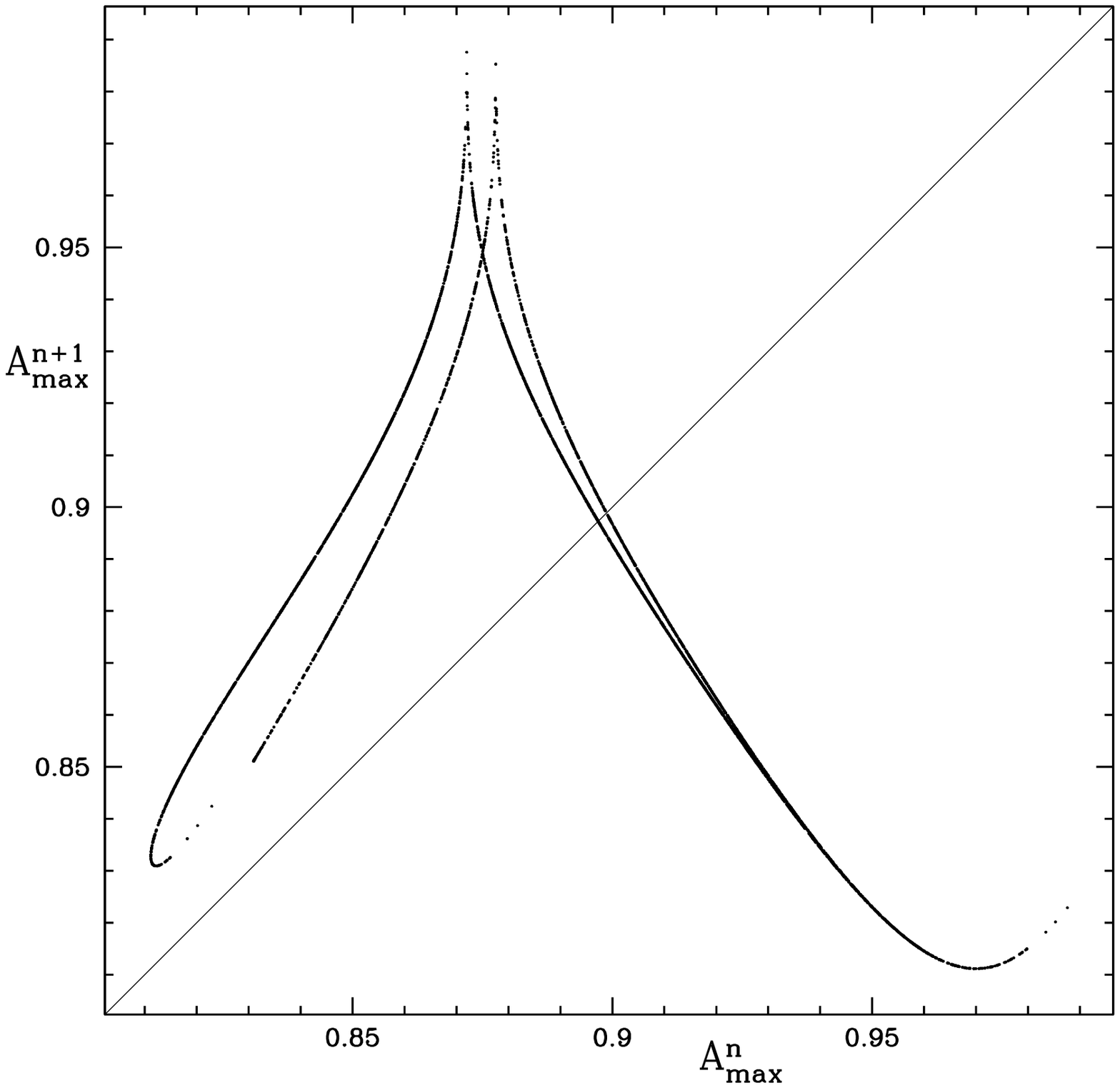}
\caption{First return  map on the successive amplitude maxima $A_m^n$ of
amplitude $A$, namely 
         $A_{max}^{n+1}$ \vs $A_{max}^n$}
\end{center}
\label{fig3}
\end{figure}
%-----------------------------------------------------------------------

%-------------------------------------------------------------

\noi 2. \underbar{The oscillatory solutions for the amplitudes}

\vskip 3pt

%These solutions exist only when both modal amplitudes are non zero, \ie both
%modes are excited.  

These solutions come in two types:

\vskip 2pt
i. Purely periodic oscillations in the amplitudes and phases, corresponding to periodic
amplitude and phase modulations of the stellar pulsations.

\vskip 2pt

ii. Irregular oscillations (irregular amplitude and phase modulations).  These
oscillations arise because of the existence of a strange attractor (chaos).  It
is these solutions that correspond to irregular Blazhko modulations.

Depending on the values of these parameters all these solutions show up in the
numerical integration of the amplitude equations.

\vskip 5pt

%-------------------------------------------------------------

The amplitude equations have too many parameters to make an exhaustive study of
their solutions possible.  However we can impose some constraints.  By placing
ourselves inside the fundamental mode instability strip we can assume that the
growth rate $\kappa_a>0$.  The quadratic terms in Eqs.~(\ref{eqs}), \viz $Q_a$,
$Q_b$, $T_a$ and $T_b$ are generally positive.
We can scale the time by setting $\kappa_a=1$ and scale the amplitudes A and B
by respectively setting $Q_a=1$ and $Q_b=1$.

%-------------------------------------------------------------

In the following we show an example of an irregular amplitude modulation.  
The values of the remaining parameters are chosen to be
$\kappa_b$ = -- 0.1, $T_a$ =
10.0, $T_b$ = 1.0, $C_a$ = 0.75, $C_b$ = 11.0, $\delta_a$ = 0.5,
$\delta_b$ = 1.2, $\Delta$ = --2.

The single mode fixed point is located at $A_0=1$ (by construction).  The
Floquet stability coefficients corresponding to an excitation in mode b
\citep{mb} $${\cal F}_b = \kappa_b -T_b A_0^2 \pm \sqrt{(C_b^2 A_0^{18}
-\Delta^2)}$$ have the values --12.034 and 9.8343, and the single mode is
therefore unstable to an excitation of mode b.

The two mode fixed point is found to be located at $A_1=0.7979$, $B_1=0.1918$,
$\Gamma_1= 0.1955$, and its stability eigenvalues are calculated to be
$\sigma_0= -3.3135$ and $\sigma_\pm - 0.2057 \pm i \thi \thi 2.7913$.  This
fixed point is therefore unstable as well and an oscillation arises in the
amplitudes.

The temporal variation of the amplitudes $A(t)$ and $B(t)$ is displayed in
Fig.~\ref{fig1}, with a blowup in the right panel.  One notes that, in this
example, the modulations in the amplitudes are irregular, akin to the observed
Blazhko cycles.

In RR~Lyr stars the relative growth rates $\kappa_0\times P_0$ are typically
of the order of a percent.  The amplitude modulations occur on time scale that
is set by $\kappa_a$ that is therefore much longer than the fundamental
pulsation period.  In our example in which we have scaled $\kappa$ to be unity,
the scaled 'Blazhko period' comes out to be $\sim 2.5$ as the figure shows,
corresponding thus to $\sim 250$ periods, in the ball park of the observed
Blazhko cycle periods.  We note though that this period can easily be shortened
or lengthened with a finetuning of the parameters in the amplitude equations and
one can get the whole range of Blazhko 'periods'.

Admittedly we have chosen the parameters in our amplitude equations in a
somewhat \adhoc way, but we find that the range of parameters over which
irregular modulations occur is quite broad.

\vskip 3pt

\section{Nature of the irregular modulations}

The reader may wonder how such irregular modulations arise.  Because of this
spiral-node nature of the fixed point, the trajectories in the complex
amplitude ($a,b$) space spiral away from the fixed point in the plane of the
spiral roots ($\sigma_\pm$), get bent around in the nonlinear regime to be
attracted back to the fixed point along the direction of the real, stable root
($\sigma_0$).  Under broad parameter conditions this results in irregular
oscillations.  This behavior is displayed in a $B(t)$ -- $A(t)$ phase plot in
Fig.~\ref{fig2} for our example.

At first sight one might think that the dynamics of this strange attractor is a
of a stretch-and-fold type similar to the well known R\"ossler band.  As a
further test, we have constructed a first return map with the maxima of
amplitude $A$, \ie a plot of the successive pairs of the $A$ amplitude,
$A_{max}^{n+1}$ \vs $A_{max}^n$, shown in Fig.~({fig3}).  It exhibits a tent
like structure more akin to that of the Lorentz attractor
\citep[\eg][]{stewart}, although somewhat more complex because of the split.
For reference we have drawn a diagonal line that intersects the map at two very
closely spaced points.  At both points the slope of the map is greater than
unity (|slope|>1) which establishes the presence of a strange attractor.

Chaotic behavior had previously been detected in the pulsations of the RV~Tauri
class star R~Scuti \citep{rscuti1,rscuti2}.  It had also been predicted to be
the cause of irregular pulsations in long period W~Vir stellar models
\citep{wvir1,wvir2,mb}.  It is interesting that the irregular modulations in
the Blazhko RR~Lyr stars would also be caused by a chaotic dynamics.  There is
however a big physical difference between the two cases.  In RR~Lyr the
amplitude modulations occur on a time scale that is long compared to the
'period' of the pulsation, because this time scale is set by the pulsational
growth rates that are small in these non adiabatic stars.  In contrast, the
W~Vir and RV~Tau stars have growth rates that are comparable to the pulsation
frequencies and therefore allow the modulations to occur over a cycle,
resulting in irregular pulsations (usually called semi-regular in the
astronomical literature).

\vskip 3pt

\section{Other resonances}

In this paper we have only considered the modal coupling between the 9th
overtone and the fundamental mode through the $9:2$ resonance.

An inspection of the period ratio diagram (Fig.~3 of \cite{KMS}) shows that the
spacings between the curves are close to 0.5, and that therefore
there is a plenitude of additional near resonances, \eg  
$$\omega_9 - \omega_7 = \omega_0 \quad {\rm and}\quad  \omega_{10} - \omega_8 =
\omega_0.$$
  
\noi Furthermore there are also several integer resonances, \eg $$\omega_7 = 4
\omega_0 \quad {\rm and} \quad \omega_5 = 3 \omega_0,$$ although this latter
resonance may not be important because of the strong damping of mode O5.  The
mode couplings that arise from these additional resonances obviously can cause
additional features and complexity in the light curves.  For the time being we
have ignored them.  Observations of RR Lyr stars indicate that the Blazhko
effect can be more complicated in some RR~Lyr stars than that explored in this
paper (\eg multi-periodic modulation as found by \cite{sodor}).  In these cases
it may necessary to include these additional resonances in the description.

\vskip 3pt

\section{Conclusions}

\cite{KMS} and \cite{setal} discovered with the help of numerical
hydrodynamical simulations that the Blazhko effect is most likely associated
with the half-integer ($9:2$) resonance between the fundamental pulsation mode
and an overtone that destabilizes the fundamental RR Lyr full amplitude
pulsation.  Because of the half integer nature of the resonance ($9P_9 = 2P_0$)
it takes therefore two fundamental periods for the pulsation to repeat, hence
the occurrence of alternating cycles.  The hydrodynamical modeling indeed
found this symmetry breaking to give rise to cycle to cycle alternations, \ie
short term variations.  The Blazhko effect, however, is a long term irregular
amplitude modulation of the pulsations which hydrodyamical simulations have not
produced so far.
 
In this paper, using an entirely different approach, namely the amplitude
equation formalism, we demonstrate that {\sl irregular} amplitude modulations
can occur quite naturally as a result of the nonlinear, resonant mode coupling
between the 9th overtone and the fundamental mode.  The phenomenon occurs over
a broad range of physical parameters and is therefore quite robust.
Furthermore we find that the range of 'periods' of the Blazhko like amplitude
modulations are in concordance with the observed ones.

It is important to emphasize that the same half integer resonance, responsible
for the destabilization of fundamental RR Lyr pulsations, is also capable of
producing period doubling or amplitude modulations, depending on the
coefficients in the equations.

However, observations may have other surprises in store and the Blazhko effect
may turn out to be more complicated, so that this simple 2 mode coupling may not
account for all its complexity. This is the case for CZ Lacertae
(\cite{sodor}), where the Blazhko-modulation is multi-periodic. There exist, in
effect, additional modal resonances or near resonances that may need to be
added in the amplitude equation description, and that may then lead to more
complicated light curves.

Finally, now that the fact that period alternations can occur in classical
variable stars such as RR~Lyr stars has been accepted by the observational
community, time might be ripe to also target BL~Her stars in which {\sl theory
had actually predicted such alternations} more than 15 years ago
\citep{blher1,blher2,blher3}, but which at the time had been received very
skeptically at best.  It is true that a recent unpublished Fourier analysis by
Buchler and Moskalik of the OGLE data of BL~Her stars does not reveal any
conclusive evidence for alternations, but then that was certainly also the case
for the pre-{\sl Kepler} of RR~Lyr.

%%%%%%%%%%%%%%%%%%%%%%%%%%%%%%%%%%%%%%%%%%%%%%%%%%%%%%%%%%%%%%%%%%%%

\acknowledgments
Fruitful discussions with R. Szab\'o are gratefully acknowledged.
This work has been supported by NSF grant AST07-07972 at UF, and by the
Hungarian OTKA grant K83790.

%%%%%%%%%%%%%%%%%%%%%%%%%%%%%%%%%%%%%%%%%%%%%%%%%%%%%%%%%%%%%%%%%%%%

\end{document}